\documentstyle[pra,preprint,aps,psfig]{revtex}

\def\be{\begin{equation}}
\def\ee{\end{equation}}
\def\ba{\begin{eqnarray}}
\def\ea{\end{eqnarray}}               

\tighten

\begin{document}

\title{Universal solutions for interacting bosons in one-dimensional
harmonic traps}

\author{Thomas Papenbrock}

\address{Physics Division, Oak Ridge National Laboratory, Oak Ridge, 
TN 37831, USA}
\maketitle

\begin{abstract}
We consider systems of interacting bosons confined to one-dimensional
harmonic traps. In the limit of perturbatively weak two-body
interactions the system exhibits several universal states that are 
exact solutions for a large class of two-body interactions.
These states are closely related to the exact solutions found
previously in rotating Bose-Einstein condensates.
\end{abstract}

\section{Introduction}

Bose-Einstein condensation in dilute atomic gases has received
considerable attention in the past few years\cite{Stringari}. A recent
experiment reports the realization of a quasi one-dimensional
Bose-Einstein condensate \cite{Druten} in a highly anisotropic,
cigar-shaped, three-dimensional trap \cite{Ketterle}. Corresponding
theoretical studies predict qualitatively different regimes for the
condensate depending on the density and the strength of the two-body
interaction. For high densities and strong repulsive interaction, the
Thomas-Fermi regime applies. Presently, this regime is accessed in the
experiment \cite{Ketterle}. For low densities and sufficiently strong
repulsive interactions, theory predicts a Girardeau-Tonks gas of
impenetrable hard cores \cite{Petrov,Dunjko,Wright}. A necessary
condition for this regime is that the interaction energy per particle
exceeds the oscillator spacing \cite{Dunjko}. If this condition is
violated, one approaches the regime of a gas with perturbatively weak
interactions.

In this work, we want to consider the regime of low densities and
perturbatively weak interactions. Within this regime the interaction
simply lifts the enormous degeneracy of the harmonically trapped
$N$-boson system. We will show that the Hilbert space and the
structure of the Hamiltonian is closely related to the problem of
rotational states in weakly interacting harmonically trapped
Bose systems in two spatial dimensions \cite{GSW,Mottelson,BP1}. This
allows us to transfer exact solutions from the latter problem to the
present case. We will present universal wave functions that are exact
solutions for a wide class of two-body interactions.

This paper is divided as follows. In the following section we state
the problem in first quantization. Next, we establish the close
relationship to rotating Bose-systems. This allows us to present
exact solutions for the one-dimensional system in analogy to the case
of rotating systems. In the fourth section we investigate the evolution
of the spectrum under change of a parameter and find the coexistence
of exact solvability and nonintegrability. We finally give a summary.

\section{Hamiltonian and Hilbert space}

Consider $N$ bosons in a one-dimensional harmonic trap that interact
via two-body contact interactions. The Hamiltonian is
\begin{equation} 
\label{ham}
H={1\over
 2}\sum_{j=1}^{N}\left(-{\partial^{2}\over \partial
 x_{j}^{2}}+x_{j}^{2}\right) +g\sum_{1\le i<j\le N}
 \delta\left(x_{i}-x_{j}\right), 
\end{equation}
where $g$ is a coupling constant.  For vanishing $g=0$ the spectrum
consists of degenerate sets of levels that differ by multiples of the
oscillator spacing.  For simplicity we set the oscillator spacing to
one and set the ground state energy to zero.  Then the energies are
simply given by integer values $E=0,1,2,\ldots$. Many-body basis
functions are completely symmetrized products
$\phi_{n_1}(x_1)\ldots\phi_{n_N}(x_N)$ of single-particle oscillator
wave functions $\phi_n(x)$, $n=0,1,2,\ldots$ and
fulfill $\sum_{j=1}^Nn_j=E$. At energy $E$ the number of degenerate
states equals the number of partitions of $E$ into at most $N$
integers: Hilbert space is partition space. Switching on the two-body
interaction lifts the degeneracy between the levels.  For
sufficiently weak interaction $N|g|\ll 1$ the quasi-degenerate levels
are much more closely spaced than the oscillator spacing, and levels
with different energies $E$ do not mix. This is the regime we are
interested in.  In first order perturbation theory, the spectrum can
be obtained by diagonalizing the interaction in the space of
degenerate levels.

We analytically treat the problem in first quantization. Let
$\hat{a}_j^\dagger$ and $\hat{a}_j$ create and annihilate an
excitation of boson $j$,
i.e. $\hat{a}_j^\dagger\,\phi_n(x_j)=\sqrt{n+1}\,\phi_{n+1}(x_j)$, and
$\hat{a}_j\,\phi_n(x_j)=\sqrt{n}\,\phi_{n-1}(x_j)$. These operators
fulfill the usual commutation relation
$[\hat{a}_i,\hat{a}_j^\dagger]=\delta_{ij}$. It is useful to express
the contact interaction in terms of creation and annihilation
operators:
\ba
\label{deriv}
\delta(x_k-x_l)
&=&\delta\left({\hat{a}_k^\dagger-\hat{a}_l^\dagger\over\sqrt{2}}
 +{\hat{a}_k-\hat{a}_l\over\sqrt{2}}\right)\nonumber\\
&=&{1\over
 2\pi}\int\limits_{-\infty}^\infty dt \,\exp{\left[i{t\over\sqrt{2}}
 \left(\hat{a}_k^\dagger-\hat{a}_l^\dagger
 +\hat{a}_k-\hat{a}_l\right)\right]}\nonumber\\
&=&{1\over
 2\pi}\int\limits_{-\infty}^\infty dt \,\exp{\left(-{t^2\over
 2}\right)} \,\exp{\left[i{t\over\sqrt{2}}\left(\hat{a}_k^\dagger
 -\hat{a}_l^\dagger\right)\right]}
 \,\exp{\left[i{t\over\sqrt{2}}\left(\hat{a}_k
 -\hat{a}_l\right)\right]}\nonumber\\
&=&{1\over
 \sqrt{2}\pi}\sum_{{m,n=0\atop m+n \,\,\,{\rm even}}}^\infty 
{\Gamma\left({m+n+1\over 2}\right)\over m !\,n !} i^{m+n}\,
 \left(\hat{a}_k^\dagger-\hat{a}_l^\dagger\right)^m\,
 \left(\hat{a}_k-\hat{a}_l\right)^n. 
\ea 
We have used the Baker-Hausdorff formula in
the step from the second to the third line in eq. (\ref{deriv}).
For perturbatively weak
interactions, only the terms in the double sum with $m=n$ are relevant
since they conserve the total number $E$ of excited
quanta. Thus,
\be
\label{contact}
\delta(x_k-x_l)={1\over
 \sqrt{2}\pi}\sum_{m=0}^\infty{\Gamma\left(m+{1\over
 2}\right)\over\left(m
 !\right)^2}(-1)^{m}\,\left(\hat{a}_k^\dagger-\hat{a}_l^\dagger\right)^m\,
\left(\hat{a}_k-\hat{a}_l\right)^m.
\ee
Note that a large class of two-body interactions which depend only on the
two-particle distance can be written similarly
to eq. (\ref{contact}). We therefore consider generalized
two-body interactions of the form
\be
\label{V}
\hat{V}=\sum_{m=0}^{\infty}c_m\hat{A}_m,
\ee
with
\be
\label{A}
\hat{A}_m=\sum_{1\le k<l\le
 N}\left(\hat{a}_k^\dagger-\hat{a}_l^\dagger\right)^m\,
\left(\hat{a}_k-\hat{a}_l\right)^m.
\ee
The coefficients $c_m$ depend on the details of the specific
two--body interaction under consideration. One finds, e.g., that a
monomial interaction of the form $(x_k-x_l)^{2n}$ yields
coefficients
\ba
c_m={(2n)!\,2^{-n}\over(n-m)!\,(m!)^2}\quad\mbox{for
 $m\le n$ and $c_m=0$ otherwise}.\nonumber
\ea
This result can be derived by
employing the characteristic function that generates the moments of
the position variable in the harmonic oscillator. Thus, zero range
interactions and interactions that are analytical in the two-particle
distance are of the general form (\ref{V}) and (\ref{A}).

The operator
\be
\label{E}
\hat{E}=\sum_k\hat{a}_k^\dagger\hat{a}_k
\ee
counts the number $E$ of excited quanta and clearly commutes with the
interaction (\ref{V}). In harmonic systems with interactions that
depend only on the two-particle distances, one may separate the motion
of the center-of-mass mode 
\be
\label{cms}
\hat{a}_c^\dagger={1\over
 N}\sum_{k=1}^N \hat{a}_k^\dagger.
\ee
The energy associated with this mode is given by
\be
\label{Ec}
\hat{E}_c=N\,\hat{a}_c^\dagger\,\hat{a}_c.
\ee
This operator commutes with the interaction (\ref{V}) and the energy
(\ref{E}). It has integer quantum numbers denoted by $E_c$, and
$E_c=0,1,2\ldots,E-2,E$ for fixed energy $E$.

\section{One-dimensional case and Yrast line problem }

When stated in terms of the operators $\hat{V}, \hat{E}$ and
$\hat{E}_c$ presented in equations (\ref{V}), (\ref{E}), and
(\ref{Ec}), respectively, the problem of interacting bosons confined
by a one-dimensional harmonic potential is very similar to the Yrast
line problem\footnote{The terminology ``Yrast line'' is borrowed from
nuclear physics.} of rotating bosons in spherically symmetric
two-dimensional harmonic traps \cite{GSW,Mottelson,BP1}. The Yrast
line problem possesses analytical solutions for several wave functions
and energies \cite{BP1,KaJa,SW,PB2,Huang,PB3,Wu,HuVo}.  We may thus
transfer these solutions to the present case. Before we do so, we
briefly remind the reader of the Yrast line problem and its results
that are relevant for us.

The Yrast line problem  consists of finding the ground states
of a rotating many-body system as a function of total angular momentum
$L$.  Let us assume that the bosons are in their ground state with
respect to excitations in the $z$-direction. The problem thus becomes
essentially two-dimensional. Again, one considers perturbatively weak
interactions that simply lift the degeneracy of the harmonic
trap. Single-particle wave functions are $\psi_l(z)=(2\pi
l!)^{-1/2}z^l$, where $z=x+iy$ is the coordinate, and we have omitted
the Gaussian in the wave function. Many-body basis states are
symmetrized products of single-particle wave functions subject to the
requirements that the single-particle angular momenta fulfill
$\sum_{j=1}^Nl_j=L$. The Hilbert space is thus partition space and
isomorphic to the Hilbert space of weakly interacting bosons in
one-dimensional traps once we identify 
$L=E$. Let us now turn to operators in the
Yrast line problem. Clearly, $z_j$ and ${\partial\over\partial{z_j}}$
are creation and annihilation operators, respectively, when acting on
the single-particle states $\psi_l(x)$. Upon substitution
\ba
\label{cor}
\hat{a}_j^\dagger&\longleftrightarrow&z_j,\nonumber\\
\hat{a}_j&\longleftrightarrow&{\partial\over\partial{z_j}},
\ea
we indeed obtain all relevant operators for the Yrast line problem
\cite{PB3}: Two-body interactions $\hat{V}$ have the form of eq. (\ref{V}),
the total angular momentum corresponds to the operator $\hat{E}$ in
eq. (\ref{E}), and the angular momentum of the center of mass is given
by the operator $\hat{E}_c$ in eq. (\ref{Ec}). Thus, there is a
correspondence between the Yrast line problem and interacting bosons
in one-dimensional harmonic traps. This correspondence is based on
the isomorphism of the Hilbert spaces and a formal identity between
operators. The Yrast line wave functions
\be
\label{Ywave}
\Psi_{L}(z_1,\ldots,z_N)=\sum_{1\le
 p_{1}<p_{2}<\ldots <p_{L}\le N} (z_{p_{1}}-z_{c})
 (z_{p_{2}}-z_{c})\ldots (z_{p_{L}}-z_{c})\, \prod_{j=1}^N{\rm
 e}^{-{1\over 2}|z_j|^2}
\ee
are exact solutions for a large class of two-body interactions for
total angular momentum $L$, and $2\le L\le N$
\cite{Huang,PB3,HuVo}. Here $z_{c}=N^{-1}\sum_{j=1}^{N}z_{j}$ denotes
the center of mass. The associated eigenenergy is
${1\over2}N(N-1)c_0+NL(c_1+2c_2)$, and the coefficients $c_0, c_1$ and
$c_2$ are determined by the specific interaction (\ref{V}) under
consideration. Let us transfer this important result to the present
case. It is a peculiarity of the Yrast line problem that the creation
operators, single-particle coordinates and single-particle wave
functions are all denoted in terms of $z_j, j=1,\ldots,N$. Clearly,
the Gaussian in the wave function (\ref{Ywave}) involves only
coordinates and no operators. Taking the correspondence between
operators (\ref{cor}), we thus find from eq. (\ref{Ywave}) that
\be
\label{wave}
\Phi_{E}(x_1,\ldots,x_N)=\sum_{1\le p_{1}<\ldots
 <p_{E}\le N} \left(\hat{a}_{p_1}^\dagger-\hat{a}_{c}^\dagger\right)
 \ldots\left(\hat{a}_{p_E}^\dagger-\hat{a}_{c}^\dagger\right)\,
\prod_{j=1}^N{\rm e}^{-{1\over 2}x_j^2}
\ee
is an eigenfunction of interaction (\ref{V}). The corresponding 
energy is \cite{PB3}
\be
\label{erg}
\epsilon(N,E)={1\over 2}N(N-1)\,c_0+NE(c_1+2c_2),
\ee
for $2\le E\le N$. This is the main result of this work.  A direct
proof of this statement may be obtained by repeating the derivation in
ref.\cite{PB3} and using the correspondence (\ref{cor}). It is based
on the observation that the state (\ref{wave}) is an eigenfunction
of the operators $\hat{A}_m$. The eigenvalues are ${1\over 2}N(N-1),
NE$ and $2NE$ for $\hat{A}_0,\hat{A}_1$ and $\hat{A}_2$, respectively.
The operators $\hat{A}_m$ with $m>2$ annihilate this state and thus 
have zero eigenvalue. By construction,
the state $\Phi_{E}$ is totally symmetric in the single-particle
coordinates, involves $E$ excited quanta, and does not excite the
center-of-mass mode (\ref{cms}), i.e. $\hat{E}_c\,\Phi_E=0$. Note that
the wave function (\ref{wave}) does not depend on the details of the
interaction (\ref{V}) since there is no reference to the coefficients
$c_j$. It is therefore a universal solution for a large class of
two-body interactions {\ref{V}). The corresponding eigenenergy (\ref{erg}),
however, does depend on the first three coefficients $c_0, c_1$ and
$c_2$.

Note that we may also establish a correspondence between basis wave
functions of the Yrast line problem and the present problem. In the
former system, wave functions are Gaussians multiplied by homogeneous
polynomials of degree $L$ that are totally symmetric in the
single-particle coordinates $z_j, j=1,\ldots,N$. Corresponding wave
functions for the present case are obtained by using the correspondence
rule (\ref{cor}) and act with the resulting operator on the Gaussian
ground state. In this basis, both wave functions (\ref{Ywave}) and
(\ref{wave}) have identical structure, i.e. identical expansion
coefficients. These wave functions are also identical in second
quantization.

We next ask the question of whether other states in the spectrum can
be written as exact algebraic functions. Leaving aside center-of-mass
excitations, there are at most two states with $E_c=0$ in the spectrum
when $E<6$ and they are both algebraic. The representation of the
second state is presented in ref.\cite{PB3}. We believe that there are
no additional universal states when $E\ge 6$, as will be seen in the
next section.

\section{Other universal solutions?}

Let us investigate whether there are further universal wave functions.
To this purpose we may take an interaction (\ref{V}) that depends on a
parameter $\tau$ and consider the spectrum as a function of the
parameter. Avoided crossings and level repulsion would certainly
indicate nonintegrability. We may further monitor the wave function
structure as a function of the parameter and thereby search for other
universal states. This requires numerical calculations.

Let us consider the interaction
\be
\label{W}
\hat{W}(\tau)\equiv(1-\tau)\,\hat{V}_{\rm 1d} + \tau\,\hat{V}_{\rm
pert}  
\ee 
which interpolates between the contact interaction $\hat{V}_{\rm 1d}$
in one-dimensional systems and the interaction $\hat{V}_{\rm pert}$ as
$\tau$ evolves from zero to one. We take $\hat{V}_{\rm
pert}=10^{-4}\,\hat{A}_4$ for definiteness.  Numerical computations are
most conveniently done in second quantization.  Let
$\hat{b}_n^{\dagger}$ and $\hat{b}_n$ create and annihilate a boson in
the single particle state $|n\rangle$ with energy $n$, respectively.
Clearly, we have $\langle x|n\rangle=\phi_{n}(x)$. At fixed energy $E$,
Hilbert space is spanned by many-body states
$|n_0,n_1,\ldots,n_E\rangle$ with $\sum_{j=0}^{E}n_{j}=N$ and
$\sum_{j=0}^{E}j\,n_{j}=E$. The matrix elements for the contact
interaction $\hat{V}_{1{\rm d}}$ involve an integral over four
oscillator functions and can be calculated analytically
\cite{Busbridge}.  The second quantized form of the operator
(\ref{contact}) reads
\begin{equation}
\label{V2}
\hat{V}_{\rm 1d}={1\over
2}\sum_{ijkl}v_{ijkl}\,\hat{b}_i^{\dagger}\,\hat{b}_j^{\dagger}\,
\hat{b}_{k}\,\hat{b}_{l},
\end{equation}
with
\ba
v_{ijkl}= {2^{-{1\over 2}}\over \pi^2}\sqrt{k!\,l!\over i!\,j!}
\sum_{t=0}^{\min{(k,l)}}
{\Gamma\left(t+{1\over 2}\right)\Gamma\left(i-t+{1\over 2}\right)
\Gamma\left(j-t+{1\over 2}\right)\over t!\,(k-t)!\,(l-t)!}\,
\delta_{i+j}^{k+l}.
\nonumber
\ea
The operator $\hat{A}_m$ has similar structure and matrix elements
\ba
a_{ijkl}^{(m)}=\sum_{\nu=\max{(0,m-k,l-i)}}^{\min{(m,l,m-i+l)}}
{m\choose\nu}{m\choose i-l+\nu}
\frac{(-1)^{i-l}\,\sqrt{i!\,j!\,k!\,l!}}
{(l-\nu)!\,(k-m+\nu)!}\,\delta_{i+j}^{k+l}. \nonumber
\ea

We compute the spectrum of $\hat{W}(\tau)$ for $0\le\tau\le 1$. We
have fixed $E=12$, $N=30$ and have restricted ourselves to the states
with $E_{c}=0$. The lowest energy is set to zero.  Fig.\ref{fig1}
shows several avoided crossings as the parameter $\tau$ evolves, and
there are no crossings. This indicates that the operator
$\hat{W}(\tau)$ is nonintegrable. Inspection of the eigenstates shows
that only the lowest energy state is independent of $\tau$. On the one
hand, this finding demonstrates the universality of the wave function
(\ref{wave}). On the other hand, it shows that there are no further
states that are universal solutions. The system of bosons in
one-dimensional harmonic traps, as well as the Yrast line problem,
thus exhibits exact solvability in coexistence with level repulsion
and nonintegrability.

Due to their nonintegrability, the systems considered in this work
differ from other exactly solvable many-body systems.  We recall that
the Calogero-Sutherland models with long-ranged inverse square
potentials are completely integrable \cite{C,S}.  Other examples
are the one-dimensional system of zero-range interacting bosons
confined to a box which is exactly solvable for all interaction
strengths \cite{Lieb} or the Tonks-Girardeau gas of hard-core bosons
in one-dimension \cite{Tonks,Girardeau}. However, the present work
shows that harmonically confined Bose systems display a few universal
wave functions that are exact solutions for a large class of
perturbatively weak two-body interactions. In this sense these fall
into the class of partially solvable quantum many-body problems
\cite{Cal}.

Let us finally compare the spectra of the contact interactions
in one-dimensional traps and the Yrast line problem. To this purpose
we choose $\hat{V}_{\rm pert}=\hat{V}_{\rm Yrast}$ and repeat the
numerical calculation.
The contact interaction $\hat{V}_{\rm Yrast}$ for the Yrast line
problem is of identical structure as the operator $\hat{V}_{\rm 1d}$
in eq. (\ref{V2}) but has matrix elements \cite{BP1}
\ba 
w_{ijkl}={(k+l)!\over 2^{k+l}\sqrt{i!\,j!\,k!\,l!}}
\,\delta_{i+j}^{k+l}. \nonumber 
\ea 
We compute the spectrum of $\hat{W}(\tau)$ for $0\le\tau\le 1$. We
choose $E=14$, $N=100$ and restrict ourselves to the states
with $E_{c}=0$. Fig.~\ref{fig2} shows the result. The lowest energy is
set to zero, and the spectra at $\tau=0$ and $\tau=1$ are scaled to
cover similar spectral ranges for display purposes.  Clearly, the
operators $\hat{V}_{\rm 1d}$ and $\hat{V}_{\rm Yrast}$ yield very
similar spectra and are in this sense close together. This suggests
that one may transfer several results from the Yrast problem to the
present case. E.g., the structure of low-lying excitations
\cite{Mottelson,KM,Ueda,Bardek,Kavoulakis,Ueda2} are expected to be
similar in both problems. Though the spectrum does not undergo
dramatic changes, we found only one universal state and a few
avoided crossings as well.

\section{Summary}

We have established a close relationship between the problem of
interacting bosons in one-dimensional harmonic traps and the Yrast
state problem for interacting bosons in two-dimensional isotropic
harmonic traps.  The Hilbert spaces of both problems are isomorphic to
each other and the important operators have identical structure.  This
allowed us to transfer several analytical results concerning
eigenstates from the Yrast problem to the one-dimensional case.  In
particular, the one-dimensional problem exhibits universal wave
functions that are exact solutions for a large class of two-body
interactions, too.  Among these are the lowest-lying excitations of
the Bose-Einstein condensate.  It is remarkable that one can thus
learn something about rotational states from wave functions in
one-dimensional, non-rotating systems. We observed level repulsion in
a parametric Hamiltonian. This indicates that the system is
nonintegrable.

The author thanks George F. Bertsch for useful discussions and a
careful reading of the manuscript. He acknowledges support as a Wigner
Fellow and staff member at Oak Ridge National Laboratory, managed by
UT-Battelle, LLC for the U.S. Department of Energy under contract
DE-AC05-00OR22725.

\begin{figure}
\centerline{\psfig{file=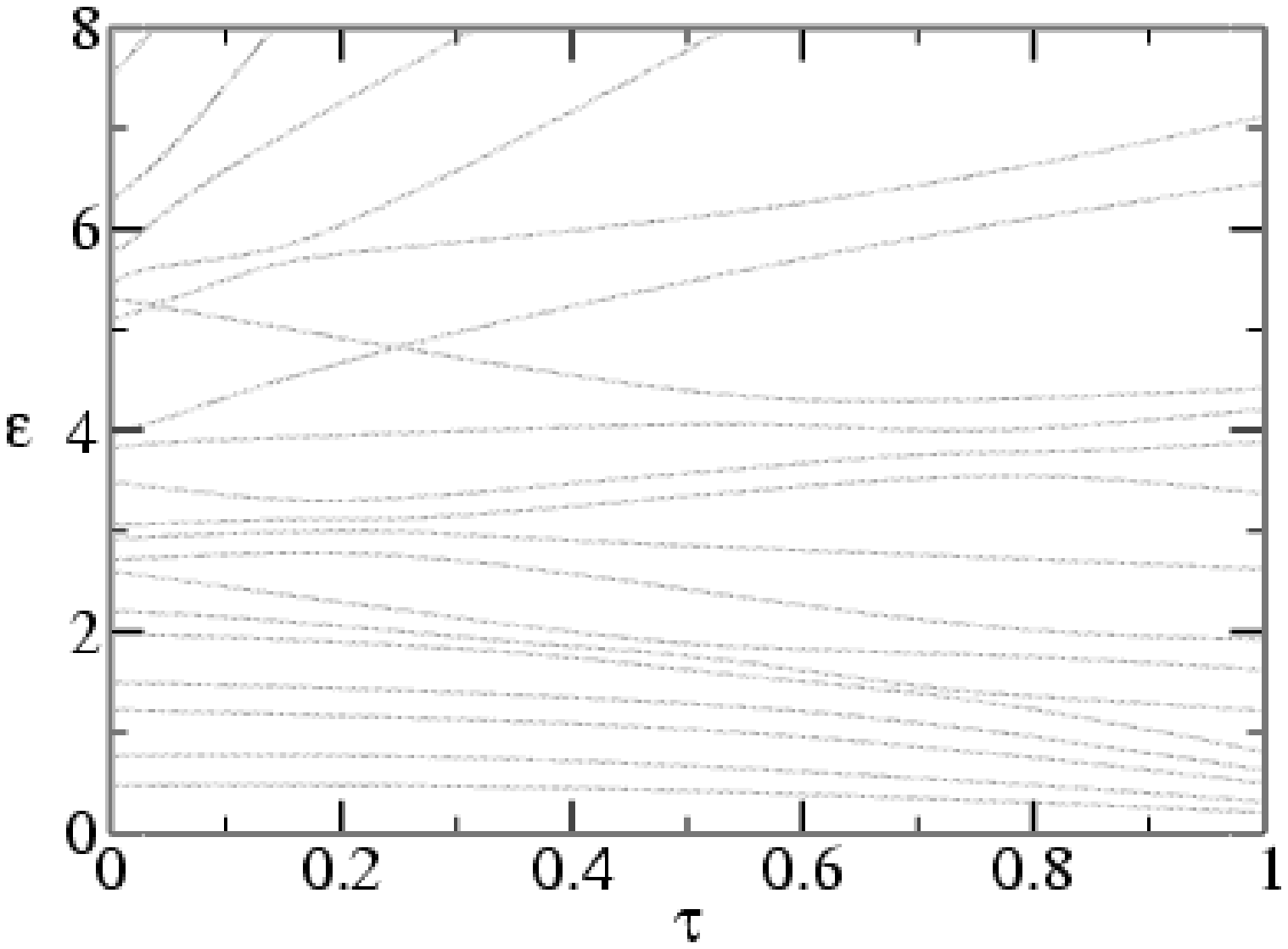,width=0.95\textwidth,angle=0}}
\protect\caption{Spectrum of 
$(1-\tau)\,\hat{V}_{\rm 1d} + 10^{-4}\,\tau\,\hat{A}_4$ 
as a function of $\tau$. There are several avoided crossings.}
\label{fig1}
\end{figure}                                                                    

\begin{figure}
\centerline{\psfig{file=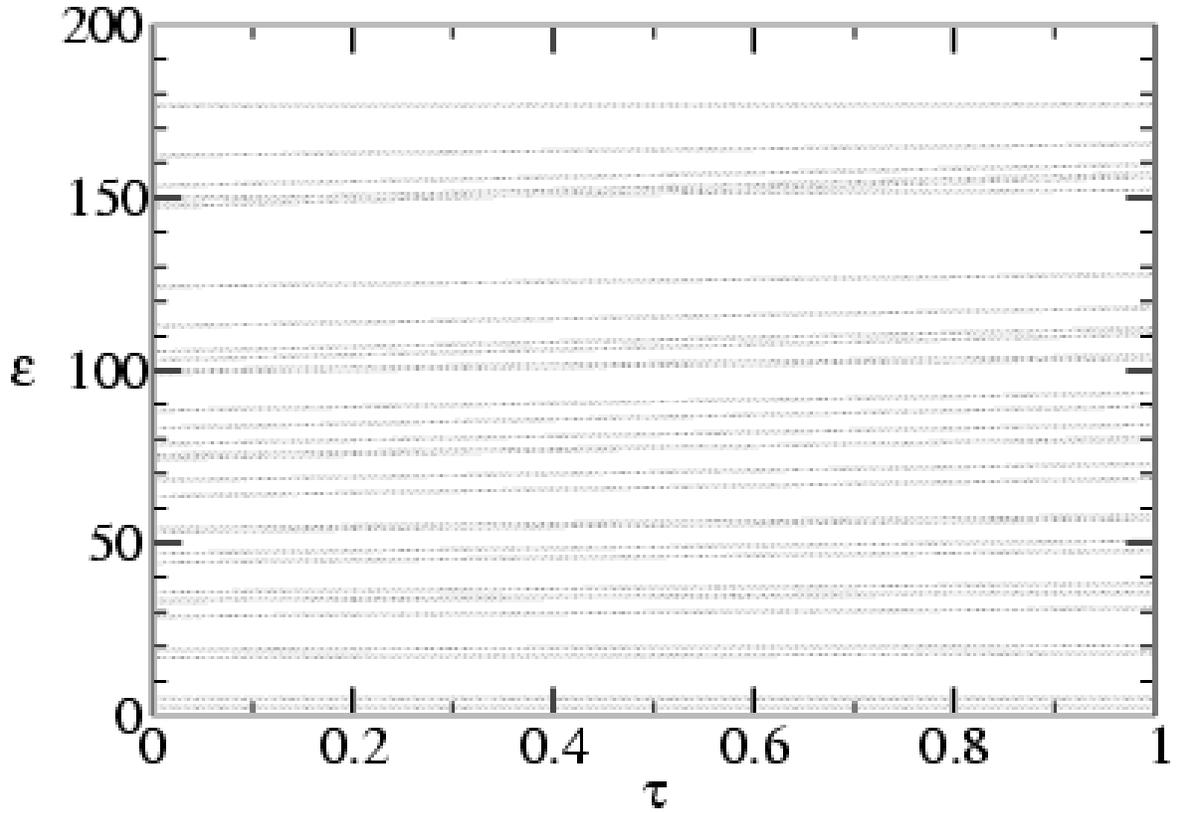,width=0.95\textwidth,angle=0}}
\protect\caption{Spectrum of 
$(1-\tau)\,\hat{V}_{\rm 1d} + \tau\,\hat{V}_{\rm Yrast}$ 
as a function of $\tau$. The spectrum exhibits only a few avoided crossings.}
\label{fig2}
\end{figure}

\end{document}